\documentclass[structabstract]{aa}

\usepackage{txfonts}
\usepackage{natbib}
\usepackage[]{graphicx}

\begin{document}

\title{An Extension of the Theory of Kinematic MHD Models of Collapsing Magnetic Traps to 2.5D with shear flow and to 3D}

\author{Keith J. Grady
\and Thomas Neukirch}

\institute{School of Mathematics and Statistics, University of St. Andrews, St. Andrews KY16 9SS, United Kingdom}

\date{ Received / Accepted}

\abstract {During solar flares a large number of charged particles are accelerated to high energies, but the exact mechanism responsible for this is, so far, still unclear. Acceleration in collapsing magnetic traps is one of the mechanisms proposed.  }{In the present paper we want to extend previous 2D models for collapsing magnetic traps to 3D models and to 2D models with shear flow.}{We use analytic solutions of the kinematic magnetohydrodynamic (MHD) equations to construct the models. Particle orbits are calculated using the guiding centre approximation.}{We present a general 
theoretical framework for constructing kinematic MHD models of collapsing magnetic traps in 3D and in 2D with
shear flow.
A few illustrative examples of collapsing trap models are presented, together with some preliminary studies of particle orbits. For these example orbits, the energy increases roughly by a factor of 5 or 6, which is consistent with the energy increase found in previous 2D models.}{}


\titlerunning{ 2.5D and 3D Collapsing Magnetic Trap Models}
\authorrunning{Grady \& Neukirch}

\maketitle

\section{Introduction}

One of the main features of solar flares is the acceleration to high energies of a substantial number of charged particles within a short period of time. The explanation of how this happens is one of the most important open questions in solar physics. There is general agreement that the energy released in solar flares is previously stored in the magnetic field, but the exact physical mechanisms by which this energy is released and converted into bulk flow energy, thermal energy, non-thermal energy and radiation energy are still a matter of discussion \citep[e.g.][]{miller:etal97,aschwanden02,neukirch05b,neukirch:etal07,krucker:etal08,aschwanden09}.
Using observations of non-thermal high-energy (hard X-ray and $\gamma$-ray) radiation, it is estimated that a large fraction of the released magnetic energy (up to  the order of 50 \%)  is converted into non-thermal energy in the form of high energy particles \citep[e.g.][]{emslie:etal04,emslie:etal05}. 

A variety of possible particle acceleration mechanisms have been suggested including 
direct acceleration in the parallel electric field associated with the reconnection process, stochastic acceleration by turbulence and/or wave-particle resonance, shock acceleration or acceleration in the inductive electric field of the reconfiguring magnetic field \citep[see e.g.][for a detailed discussion and further references]{miller:etal97,aschwanden02,neukirch05b,neukirch:etal07,krucker:etal08}. So far, none of the proposed mechanisms can explain the high-energy particle fluxes within the framework of the standard solar flare thick target model. This has recently prompted suggestions of alternative acceleration scenarios \citep[e.g.][]{fletcher:hudson08,birn:etal09}.

\citet{somov92} and \citet{somov:kosugi97} suggested that the reconfiguration of the magnetic field during a flare could contribute to the acceleration of particles. Due to the geometry of the magnetic field charged particles could be trapped while the magnetic field lines relax dynamically. In such a collapsing magnetic trap (CMT from now on) the kinetic energy of the particles could increase due to the betatron effect, as the magnetic field strength in the CMT increases, and due to first-order Fermi acceleration, as the distance between the mirror points of particle orbits decreases due to the shortening of the field lines.
There is also some observational evidence of post-flare field lines relaxation (field line shrinkage) from Yohkoh \citep[e.g.][]{forbes:acton96} and Hinode \citep[e.g.][]{reeves:etal08a} observations.

Various fundamental properties of the particle acceleration process in CMTs have been investigated by Somov and co-workers \citep[e.g.][]{bogachev:somov01,bogachev:somov05,bogachev:somov09,kovalev:somov02,kovalev:somov03a,kovalev:somov03b,somov:bogachev03}, including the relative efficiencies of betatron and Fermi acceleration,  the effect of collisions, the role of velocity anisotropies  and the evolution of the energy distribution function in a CMT. In all cases a basic model for CMTs has been used. \citet{karlicky:kosugi04} also investigated particle acceleration, plasma heating and the resulting X-ray emission using a simple CMT model and a simplified equation of motion for the particles. \citet{karlicky:barta06} used CMT-like electromagnetic fields taken from an MHD simulation of a reconnecting current sheet to investigate acceleration using test particle calculations with a view to explain hard X-ray loop-top sources.  Some indication that CMTs might be relevant for X-ray loop top sources has been provided by \citet{veronig06}. A very simple time-dependent trap model was also used by
\citet{aschwanden04} to explain the pulsed time profile of energetic particle injection during flares.

A general theoretical framework for more detailed analytical CMT models based on kinematic MHD, i.e. with given bulk flow profile, in Cartesian coordinates was presented by \citet{giuliani:etal05}  for 2D and 2.5D magnetic fields, but excluding flow in the invariant direction. Some examples of model CMTs were given together with a calculation of a particle orbit based on non-relativistic guiding centre theory \citep[see e.g.][]{northrop63}. It was found that in the models studied the curvature drift and the 
gradient-$B$ drift play an important role in the acceleration process. Similar findings have also been made, albeit in systems of a much smaller length, in the investigation of particle acceleration in particle-in-cell simulations of collisionless magnetic reconnection \citep[e.g.][]{hoshino:etal01}.

The advantage of kinematic MHD models compared to e.g. MHD simulations is that they allow us to obtain analytical expressions for the electromagnetic fields of the CMT. This makes the integration of particle orbits more accurate, because there is no need for interpolation of the fields between grid points. Furthermore the investigation of different model features is possible in an easy way by varying model parameters. The major disadvantage of kinematic MHD models is their lack of self-consistency, but this is not too critical for the purpose of test particle calculations.

Particle acceleration through rapid reconfiguration of the magnetic field has also been identified as one of the mechanisms for particle energization during magnetospheric substorms 
\citep[e.g.][]{birn:etal97,birn:etal98,birn:etal04}. During a substorm the stretched magnetic field of the magnetotail reconnects, leading to a so-called dipolarisation of the near-Earth tail, which is in principle very similar to the evolution of the magnetic field in a CMT associated with a solar flare. A general comparison of flare and substorm/magnetotail phenomena based on observations has recently been presented by \citet{reeves:etal08b}.

The purpose of the present paper is to extend the theoretical framework for kinematic MHD CMT models given by \citet{giuliani:etal05} to 2.5D models with flow in the invariant direction and to fully three-dimensional models. This is necessary for a number of reasons:

\begin{enumerate}

\item The theory of kinematic MHD CMTs as developed so far by \citet{giuliani:etal05} only allows for a magnetic field component in the invariant direction, but not for a component of the flow velocity in this direction. Without this component of the flow velocity the magnetic field component in the invariant direction can only increase in a CMT due to magnetic flux conservation.
It is, however, to be expected that during a flare magnetic shear will be reduced rather than increased and therefore the introduction of a component of the flow velocity in the invariant direction is a necessary extension to be able to make the 2.5D models more realistic.

\item In the 2D cases investigated by \citet{giuliani:etal05} the acceleration due to curvature and gradient-B drift occurs in the invariant direction. This is due to the fact that the particles gain energy while moving parallel or anti-parallel (in the case of electrons) to the inductive electric field which in a 2D trap is in the invariant direction. Due to the spatial symmetry the electric field does not vary in this direction and this will have an influence on the acceleration process.
It is therefore important to investigate the differences of the acceleration process between 2D models and non-symmetric 3D CMT models in the future.

\item \citet{giuliani:etal05} have already discussed a possible way of extending the 2D theory to three dimensions using Euler potentials. While Euler potentials allow a relatively straightforward extension of the  theory to 3D by simple analogy to the 2D case, they are not easy to use in the modelling process, which is already intrinsically more difficult in three dimensions. We therefore present in this paper an extension to the theory which makes it possible to avoid the explicit calculation of Euler potentials and uses the magnetic field directly.

\end{enumerate}

The paper is organised as follows. In Sect. \ref{sec:2d_theory} we briefly summarise the present state of the kinematic MHD theory of CMTs, before presenting its extensions to 2.5D with shear flow and to 3D.
In Sect. \ref{sec:examples} a couple of  illustrative examples of CMT models based on the new theoretical descriptions are shown, followed by examples of test particle calculations in Sect. \ref{sec:orbits}. We conclude the paper in Sect. \ref{sec:summary} with a summary and conclusions. Appendix A gives more detail of the calculation of the 3D field using Euler Potentials.

\section{Basic Theory}\label{sec:theory}

The CMT is assumed to form outside the nonideal reconnection region, so the ideal kinematic MHD equations may be used to describe the evolution of the electromagnetic field 
\begin{eqnarray}
  \vec{E} + \vec{v} \times \vec{B} & = & \vec{0}, \label{eq:ohms}\\
  \frac{\partial \vec{B}}{\partial t} & = & - \nabla \times \vec{E}, \label{eq:faraday}\\
  \nabla \cdot \vec{B} & = & 0, \label{eq:divb}
\end{eqnarray}
with the MHD velocity $\vec{v}$ assumed to be given as a function of space and time. We will also make occasional use of the ideal induction equation
\begin{equation}
\frac{\partial \vec{B}}{\partial t}  =   \nabla \times (\vec{v} \times \vec{B}),
\label{eq:induction}
\end{equation}
which results from combining Eqs. (\ref{eq:ohms}) and (\ref{eq:faraday}).

\subsection{Kinematic MHD Models of CMTs in 2.5D without shear flow}
\label{sec:2d_theory}

We start by giving a brief overview of the  translationally invariant 2.5D kinematic MHD theory of CMTs developed by \citet{giuliani:etal05}. This does not include a velocity component in the invariant direction. 
In the following we will use the same coordinate system as used by  \citet{giuliani:etal05}, i.e. all physical quantities depend only upon $x$ and $y$, with $x$ being the coordinate parallel to the solar surface (photosphere) and $y$ being the height above the solar surface. The invariant direction is the $z$-direction.

For the cases with spatial symmetry it is useful to write the magnetic field as
\begin{equation}
\vec{B} = \vec{B}_p +B_z \vec{e}_z=\nabla A \times \vec{e}_z + B_z \vec{e}_z ,
\label{eq:2DB_def}
\end{equation}
where $A(x,y,t)$ is the flux function, $\vec{B}_p=(B_x(x,y,t),B_y(x,y,t),0)$ and $B_z(x,y,t)$ the $z$-component of the magnetic field.
An important assumption made by \citet{giuliani:etal05} is that there should be no flow in the invariant direction, i.e.
\begin{equation}
\vec{v}_2(x,y,t) = (v_x(x,y,t) ,v_y(x,y,t),0).
\label{velocity2d}
\end{equation}
As we will make use of this particular velocity field later on, we use the index $2$ to distinguish it from the full velocity field with non-zero $v_z$.
Using an appropriate gauge for $A$, the $z$ component of Ohm's law (\ref{eq:ohms}) gives
\begin{equation}
\frac{dA}{dt} = \frac{\partial A}{\partial t} + \vec{v}_2 \cdot \nabla A = 0
\label{eq:2d_prop}
\end{equation}
for the time evolution of the flux function $A$.
For the time evolution of $B_z$ it is better to use the $z$-component of the induction equation (\ref{eq:induction}),
\begin{equation}
%Induction eq
\frac{{\partial B_z }}{{\partial t}} + \nabla  \cdot ({\vec{v}_2}B_z ) = 0. 
\label{eq:2d_ind}
\end{equation}
Equations (\ref{eq:2d_prop}) and (\ref{eq:2d_ind}) simply express the conservation of magnetic flux. In the case with vanishing shear velocity ($v_z=0$) the  magnetic flux $\int B_z dx dy$ is conserved independently.
To solve Eqs. (\ref{eq:2d_prop}) and (\ref{eq:2d_ind}) for $A(x,y,t)$ and $B_z(x,y,t)$, \citet{giuliani:etal05} prescribe a time-dependent transformation between Lagrangian coordinates $X$, $Y$ and Eulerian coordinates $x$, $y$: 
\begin{equation}
X=X(x,y,t),\qquad Y=Y(x,y,t) ,
\end{equation}
instead of a time-dependent velocity field $v_x(x,y,t)$, $v_y(x,y,t)$. The velocity field can be determined easily from the transformation equations (see Eqs. (23)-(26) of \citet{giuliani:etal05}).

The solution for the magnetic flux function $A(x,y,t)$ is then trivially given by 
\begin{equation}
A(x,y,t) = A_0(X(x,y,t),Y(x,y,t))\mbox{,}
\label{eq:A2D_sol}
\end{equation}
where $A_0(X,Y)$ is the flux function at some reference time $t=t_0$. The $B_x$- and $B_y$-components of the magnetic field can be calculated from Eq. (\ref{eq:2DB_def}) by differentiation.

Equation (\ref{eq:2d_ind}) has the form of a continuity equation for $B_z$ with the solution
\begin{equation}
B_z(x,y,t)= J^{-1} B_{0z}(X(x,y,t),Y(x,y,t))\mbox{,}
\label{eq:Bz2Dnoshear_sol}
\end{equation}
where $B_{0z}(X,Y)$ is again the $B_z$ at a reference time $t=t_0$ and $|J|$ is the Jacobian determinant of the transformation between the Lagrangian and Eulerian coordinates, here written as
\begin{equation}
J^{-1} = \frac{\partial X}{\partial x} \frac{\partial Y}{\partial y} - 
 \frac{\partial Y}{\partial x} \frac{\partial X}{\partial y} . 
\end{equation}
The Jacobian determinant basically expresses the deformation of infinitesimal area elements , i.e. a change of cross section, 
in the $x$-$y$-plane during the time evolution of the system. Because the magnetic flux associated with $B_z$ is conserved independently in the case discussed in this section, any decrease in area must be compensated by a matching increase in $B_z$ and vice versa.

Finally, the electric field can be determined from Ohm's law (\ref{eq:ohms}) once the velocity field 
$\vec{v}$ and the magnetic field $\vec{B}$ are known.

\subsection{Extension to 2.5D with shear flow}

To allow the effect of shearing and also de-shearing of the magnetic field to be taken into account it is necessary to 
have a non-zero $v_z(x,y,t)$. The basic effect of a non-zero $v_z$ is to add a source term to equation (\ref{eq:2d_ind}) 
\begin{equation}
\frac{\partial B_z}{\partial t} + \nabla \cdot \left( \vec{v}_2 B_z \right) = \nabla \cdot \left( v_z \vec{B}_p\right) .
\label{eq:2d_ind_shear}
\end{equation}
The source term on the right-hand-side of Eq. (\ref{eq:2d_ind_shear}) destroys the separate conservation of magnetic flux in the $z$-direction, because a non-zero $v_z$ allows $B_x$ and $B_y$ to by turned into $B_z$ and vice versa.
In addition to the transformation equations for the $x$- and $y$-coordinates one has to add a transformation equation for the $z$-coordinate of the form
\begin{equation}
Z= z+ \bar{Z}(x,y,t).
\label{eq:ztrans}
\end{equation}
The general solution for the flux function remains the same, but the solution for $B_z$ becomes more complicated. As it is much easier to deduce the solution for $B_z$ as a special case from the 3D case discussed next, we will give the appropriate expressions for $B_z$ and the velocity field after discussing
the general theory for three dimensions.

\subsection{Extension to 3D}

As already pointed out by \citet{giuliani:etal05}, one can in principle use a similar approach as for 2D to generalise the theory to 3D. Instead of writing the magnetic field in terms of a flux function $A$ we use 
Euler Potentials  to satisfy the solenoidal condition (\ref{eq:divb}) 
\citep[see e.g.][]{stern70,stern87}:
\begin{equation}
\vec{B}=\nabla \alpha \times \nabla \beta .
\label{eq:euler}
\end{equation}
When using Euler potentials one has to assume that the magnetic topology of the CMT is sufficiently simple to allow the global
 existence of a set of Euler potentials satisfying Eq. (\ref{eq:euler}) for all positions and times \citep[see e.g.][for a discussion]{moffatt78}.
Using Euler potentials in an appropriate gauge, Ohm's law (\ref{eq:ohms}) can be written as \citep[e.g.][]{stern70}
\begin{eqnarray}
\frac{\partial \alpha}{\partial t} + \vec{v} \cdot \nabla \alpha &=& 0, \label{eq:alpha}\\
\frac{\partial \beta}{\partial t} + \vec{v} \cdot \nabla \beta &=& 0, \label{eq:beta}
\end{eqnarray}
and the solutions of Eqs. (\ref{eq:alpha}) and (\ref{eq:beta}) are given by
\begin{eqnarray}
  \alpha \left( \vec{x},t \right) & = & \bar{\alpha} \left( \vec{X} \! \left( \vec{x},t \right) \right), \label{eq:solalpha} \\
  \beta  \left( \vec{x},t \right) & = & \bar{\beta} \left( \vec{X} \! \left( \vec{x},t \right) \right) , \label{eq:solbeta}
\end{eqnarray}
where, as in the 2D solution $\bar{\alpha} \left( \vec{X} \right)$ and $\bar{\beta} \! \left( \vec{X} \right)$ are 
the Euler potentials at a reference time $t=t_0$.\footnote{For example, \citet{giuliani:etal05} use the final time as reference time.}
As in the 2D case a transformation between Eulerian ($\vec{x}$) coordinates and Lagrangian ($\vec{X}$) 
coordinates is assumed as given in the form
\begin{equation}
  \vec{X} = \vec{X}\left(x,y,z,t\right)\mbox{,} \label{eq:3dtrans}
\end{equation}
where we have combined the transformation equations for the three coordinates ($X=X(x,y,z,t)$, 
$Y=Y(x,y,z,t)$, $Z=Z(x,y,z,t)$)
into a vector $\vec{X}=(X,Y,Z)$ for ease of reference. For completeness, the full derivation of the expression for the magnetic field using Eqs. (\ref{eq:solalpha}) and (\ref{eq:solbeta}) is 
shown in Appendix \ref{appendixa}. The result is given by the equations
\begin{eqnarray}
      B_x &=& 
          \left( \frac{\partial{}\vec{X}}{\partial{}y} \times \frac{\partial{}\vec{X}}{\partial{}z} \right) \cdot
           \vec{B}_0 \left( \vec{X} \right) \mbox{,}
           \label{eq:Bx3d}  \\
           B_y&=& \left( \frac{\partial{}\vec{X}}{\partial{}z} \times \frac{\partial{}\vec{X}}{\partial{}x} \right) \cdot
           \vec{B}_0 \left( \vec{X} \right) \mbox{,}
           \label{eq:By3d} \\
      B_z&=& \left( \frac{\partial{}\vec{X}}{\partial{}x} \times \frac{\partial{}\vec{X}}{\partial{}y} \right) \cdot
           \vec{B}_0 \left( \vec{X} \right) \mbox{.}          
\label{eq:Bz3d}
\end{eqnarray}
It is important to note that this result is expressed completely in terms of derivatives of the transformation equations and the magnetic field at the reference time $t=t_0$
\begin{equation}
  \vec{B}_0 = \vec{B}_0\left(x,y,z\right), \label{eq:B0}
\end{equation}
 i.e. no reference to Euler potentials has to be made when modelling CMTs in 3D. This is no surprise as the same result can also be found without the use of Euler potentials \citep[see e.g.][p. 44]{moffatt78}, but using Euler potentials makes the transition from 2D to 3D a bit more obvious. While Euler potentials are often very useful for gaining better theoretical insight \citep[e.g.][]{stern70,hesse:schindler88,hesse:etal05}, they are usually quite difficult to use for modelling purposes \citep[e.g.][]{platt:neukirch94,romeou:neukirch99,romeou:neukirch02}.
Also, due to this result the conditions for the global existence of Euler potentials do not apply for the modelling of 3D CMTs and the modelling process is thus much less restrictive. It is therefore very beneficial to have a formulation which is based purely on the magnetic field at the reference time 
and on the transformation equation (\ref{eq:3dtrans}), both of which we are free to choose.

From the transformation equation (\ref{eq:3dtrans}), one can calculate the flow velocity by using that
\begin{equation}
\frac{d \vec{X}}{d t} = \frac{\partial \vec{X}}{\partial t} + \left( \vec{v} \cdot \nabla \right) \vec{X} =\vec{0} ,\label{eq:3d_flow}
\end{equation}
from which one can calculate the velocity $\vec{v}$ by inversion of the non-singular $3\times3$ matrix 
$\nabla \vec{X}$, giving
\begin{equation}
\vec{v} (x,y,z,t) =- (\nabla \vec{X})^{-1} \cdot \frac{\partial \vec{X}}{\partial t}.
\label{eq:3d_velocity}
\end{equation}
We refrain from giving the complete explicit form of the velocity field here, as it is rather lengthy and not too instructive.
Finally, knowledge of the flow velocity and the magnetic field allows the calculation of the electric field from Ohm's law (\ref{eq:ohms}).

\subsection{Derivation of the 2.5D case with shear flow formulae from the 3D case}

We will now come back to the 2.5D case with shear flow.
The transformation equation (\ref{eq:ztrans}) for the $z$-coordinate implies that
\begin{equation}
 \frac{\partial{}\vec{X}}{\partial{}z} = \left( 0,0,1 \right).
 \label{eq:diffxx-z}
\end{equation}
Using Eq. (\ref{eq:Bz3d}), the $z$-component of the magnetic field for the 2.5D case with shear  is given by
\begin{eqnarray}
B_z(x,y,t) &=&  \left( \frac{\partial Y}{\partial x} \frac{\partial \bar{Z}}{\partial y} - 
\frac{\partial \bar{Z}}{\partial x} \frac{\partial Y}{\partial y} \right) B_{0x} \left( \vec{X} \right) \nonumber \\
& & +\left( \frac{\partial \bar{Z}}{\partial x} \frac{\partial X}{\partial y} -
 \frac{\partial X}{\partial x} \frac{\partial \bar{Z}}{\partial y} \right)B_{0y} \left( \vec{X} \right) \nonumber \\
& & + \left( \frac{\partial X}{\partial x} \frac{\partial Y}{\partial y} - 
 \frac{\partial Y}{\partial x} \frac{\partial X}{\partial y} \right)B_{0z} \left( \vec{X} \right).
\label{eq:Bzsol_shear}
\end{eqnarray}
The last term of Eq. (\ref{eq:Bzsol_shear}) is identical to the 2.5D solution for $B_z$ without shear flow given in Eq. (\ref{eq:Bz2Dnoshear_sol}). The other two terms represent the extra possibility of 
turning $B_x$ or $B_y$ flux into $B_z$ flux and vice versa.

The velocity field can be determined by using the transformation equations for the 2.5D case with shear flow in Eq. (\ref{eq:3d_flow}).
This gives the components of the velocity as
\begin{eqnarray}
v_x&=&\left(-\frac{\partial X}{\partial t} \frac{\partial Y}{\partial y} +
      \frac{\partial X}{\partial y} \frac{\partial Y}{\partial t} \right)
     \left(
      \frac{\partial X}{\partial x} \frac{\partial Y}{\partial y}
      -\frac{\partial Y}{\partial x} \frac{\partial X}{\partial y} \right)^{-1} \mbox{,}\\
v_y&=&
\left(-\frac{\partial X}{\partial x} \frac{\partial Y}{\partial t} +
       \frac{\partial X}{\partial t}\frac{\partial Y}{\partial x} \right)  
       \left(
      \frac{\partial X}{\partial x} \frac{\partial Y}{\partial y}
      -\frac{\partial Y}{\partial x} \frac{\partial X}{\partial y} \right)^{-1} \mbox{and} 
\end{eqnarray}
\begin{eqnarray}
v_z&=& -\frac{\partial \bar{Z}}{\partial t} - 
            \left[\frac{\partial \bar{Z}}{\partial x} \left(- \frac{\partial X}{\partial t}\frac{\partial Y}{\partial y} +
           \frac{\partial X}{\partial y} \frac{\partial Y}{\partial t} \right) \right. \nonumber \\
    & & +  \left. \frac{\partial \bar{Z}}{\partial y} 
        \left(-\frac{\partial X}{\partial x} \frac{\partial Y}{\partial t} +
           \frac{\partial X}{\partial t} \frac{\partial Y}{\partial x}  \right)\right] 
      \left(
      \frac{\partial X}{\partial x} \frac{\partial Y}{\partial y}
      -\frac{\partial Y}{\partial x} \frac{\partial X}{\partial y} \right)^{-1} \mbox{.}
\end{eqnarray}
Again, the electric field can be calculated from Ohm's law (\ref{eq:ohms}), once the velocity field and the magnetic field are known, but due to the complexity of the expressions we do not state them here 
explicitly.

\section{Illustrative Examples of Collapsing Trap Models}
\label{sec:examples}

In the following we shall discuss some simple illustrative examples of CMTs in 2.5D with shear flow 
and in 3D. Our main purpose here is to compare some of the features of these extended models with the results found by \citet{giuliani:etal05} for 2D models. Therefore, we shall use one of the transformations used by \citet{giuliani:etal05}. We do not suggest that these examples can be regarded as realistic models of a flare, but they offer some insight into the basic features of 2D and 3D collapsing trap models.

\subsection{An illustrative example for a 2.5D CMT model with shear flow}
\label{sec:2d}

We first  add a shear flow to the main example presented in \citet{giuliani:etal05}. Therefore, as in their paper, the 2D magnetic field is generated using the flux function
\begin{equation}
A_0=c_1 \arctan \left( \frac{y_0 + d/L}{x_0 + 1/2}\right) - c_1 \arctan \left( \frac{y_0 + d/L}{x_0 - 1/2}\right),
\label{AGiuliani}
\end{equation} 
which represents a loop between two line currents at $x_0=\pm L/2$, i.e. separated by a distance $L$ and placed at a distance $y_0=-d$ below the photosphere. The magnetic field generated by the flux function (\ref{AGiuliani}) is potential if regarded as a function of $x_0$ and $y_0$. This potential field
is the final field to which the CMT relaxes as $t \to \infty$. In the model presented in this section, the magnetic field in the $z$-direction is set to zero, $B_z=0$, as $t\to \infty$.

At other times, the magnetic field will be non-potential and we will choose a coordinate transformation which gives an initially sheared magnetic field, i.e. with $B_z \ne 0$.
To ensure continuity from the model of \citet{giuliani:etal05} to our model the transformations of the $x$- and $y$-coordinates are the same as in their paper, i.e.
\begin{eqnarray}
x_0&=&x \mbox{,} \\
y_0&=&(a t)^b \ln \left[ 1+ \frac{y}{(a t)^b}\right]
 \left\{ \frac{1+ \tanh[(y- L_v/L) a_1]}{2}\right\} \nonumber \\
& & +\left\{ \frac{1+ \tanh[(y- L_v/L) a_1]}{2}\right\} y \mbox{.}
 \label{ytransform2d}
\end{eqnarray} 
This transformation basically stretches the  magnetic field in the $y$-direction above a height 
given by $L_v/L$, where the transition between unstretched and stretched field is controlled by the parameter $a_1$. We use the same parameter values as \citet{giuliani:etal05}, namely $a=0.4$, $b=1.0$, $L_v/L=1$ 
and $a_1=0.9$. 
For simplicity, the transformation depends on time only through the function $y_0(y,t)$. This time-dependence lets the field collapse to the final field described above as for $t\to\infty$, $y_0$ tends to $y$.
Other important features of the transformation are that the foot points of magnetic field lines do not move during the collapse as for $y=0$ we have $y_0=0$ for all $t$.

The important difference to the model used by \citet{giuliani:etal05} is that we introduce an additional
transformation for the $z$-coordinate giving rise to
a shearing flow as discussed above. The transformation for the $z$-coordinate is chosen as
\begin{equation}
z_0=z+\delta \left[ y_0(y,t) -y \right] \frac{x}{a^2_{\rm{2.5D}}+x^2},
\label{ztransform2d}
\end{equation}
where $\delta$ and $a_{\rm{2.5D}}$ are parameters that are explained later.
This transformation induces an $x$- and $y$-dependent shear motion. The reasoning behind choosing
the transformation as given is as follows:

\begin{enumerate}

\item the shear flow should be anti-symmetric with respect to $x$ and vanish as $|x| \to \infty$, which is achieved in a simple way by the $x$-dependence of the transformation  and controlled by the parameter $a_{2.5D}$;

\item the shear flow should vanish at the photosphere (no foot point motion) and be of 
noticeable strength only in the stretched area of the magnetic field and it should also vanish as $t\to \infty$; this is achieved in a simple way by the $y$-dependence of the transformation;

\item we should be able to control the magnitude of the shear flow, 
which is done by the parameter $\delta$.

\end{enumerate}

\begin{figure}
\resizebox{\hsize}{!}{\includegraphics[clip]{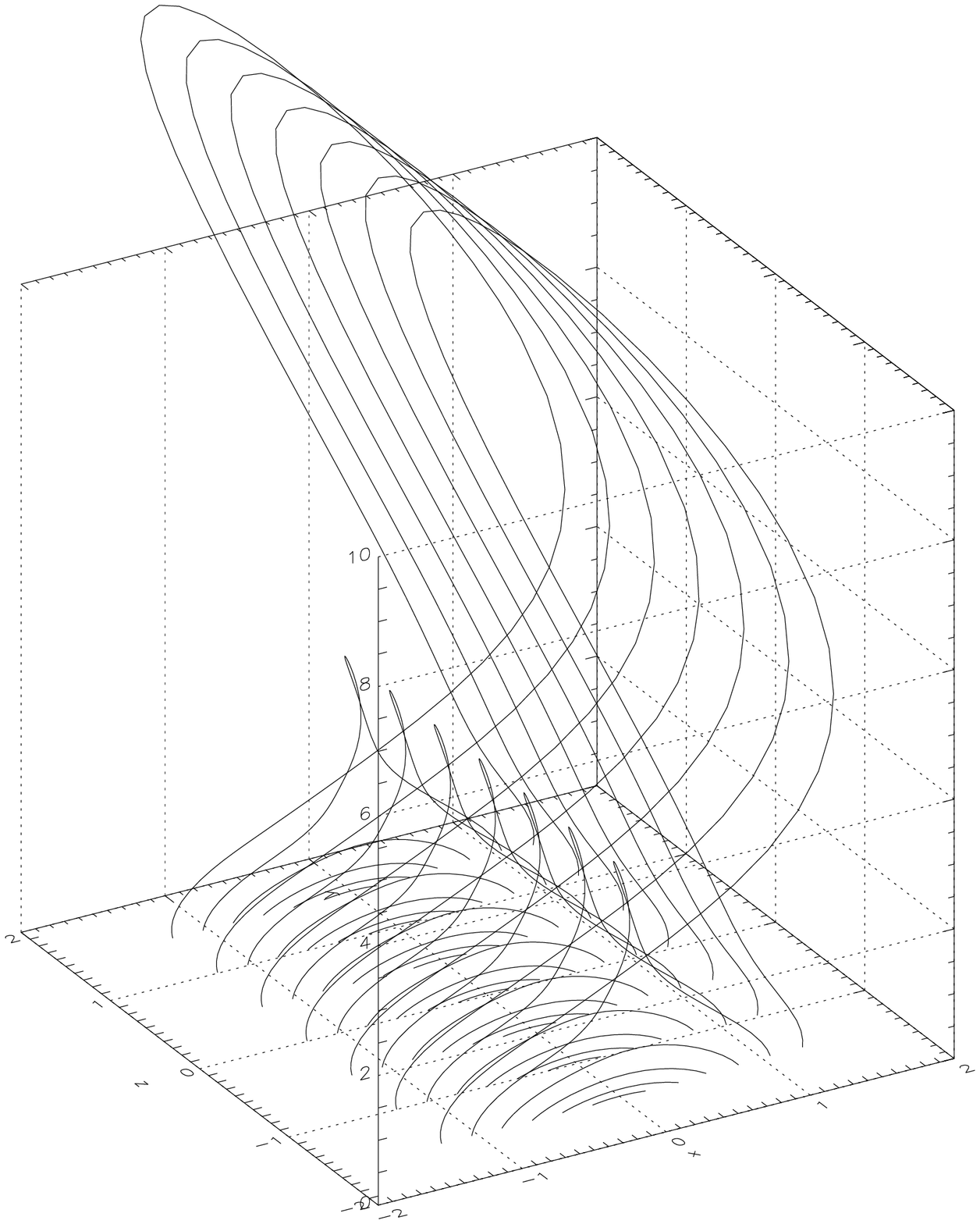}\includegraphics[clip]{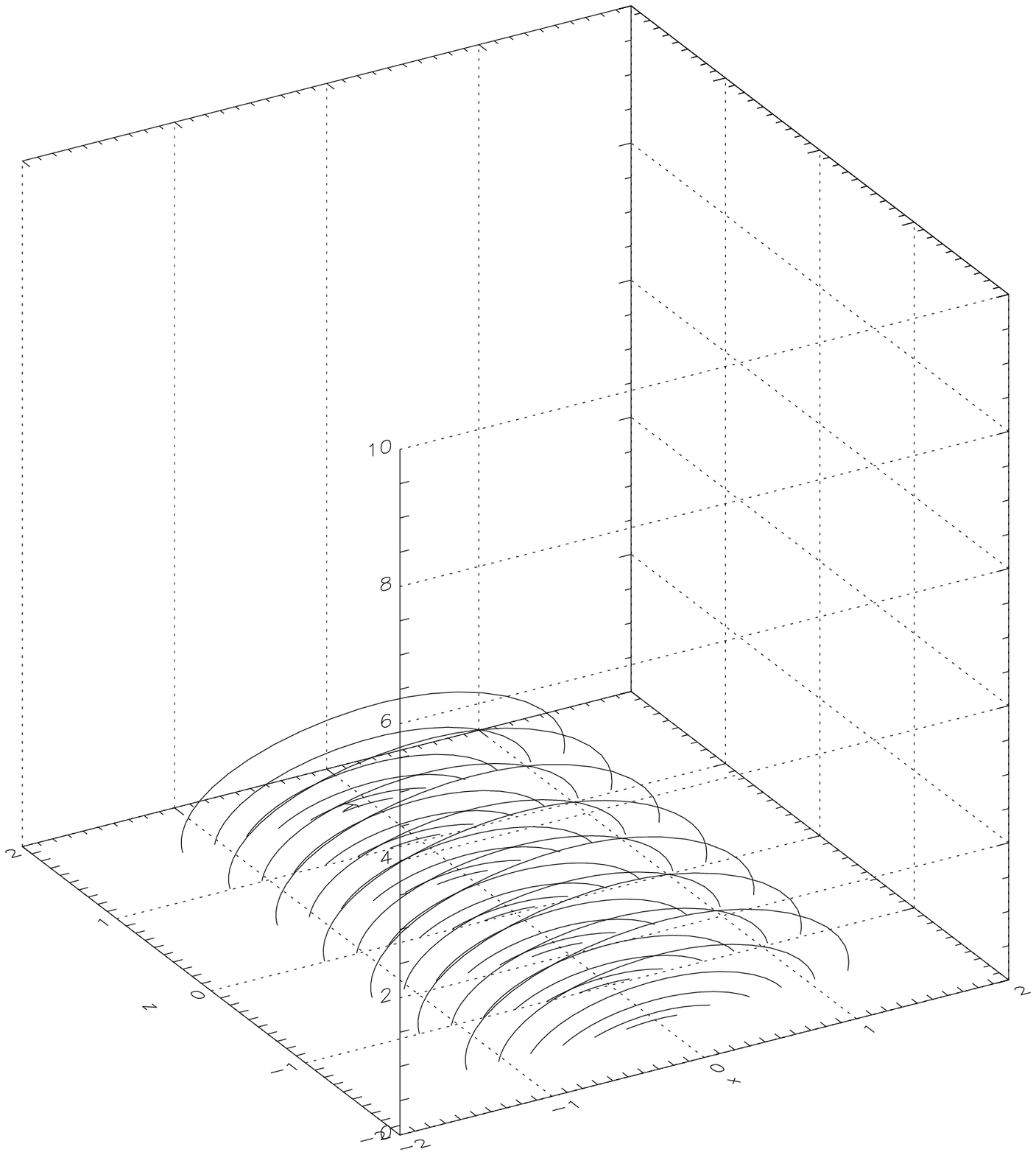}}
\caption{Field lines in the example 2D case with shear flow. Lengths are normalised to $L=10Mm$. The left plot shows the magnetic field at $1.05s$, the right plot shows it at $50.8s$. The collapse of the field lines in the $y$-direction is obvious. Note the difference in scale between the $x$-$z$-plane and the 
$y$-direction.}
\label{fig:2d_field}
\end{figure}

\begin{figure}
\resizebox{\hsize}{!}{\includegraphics[clip]{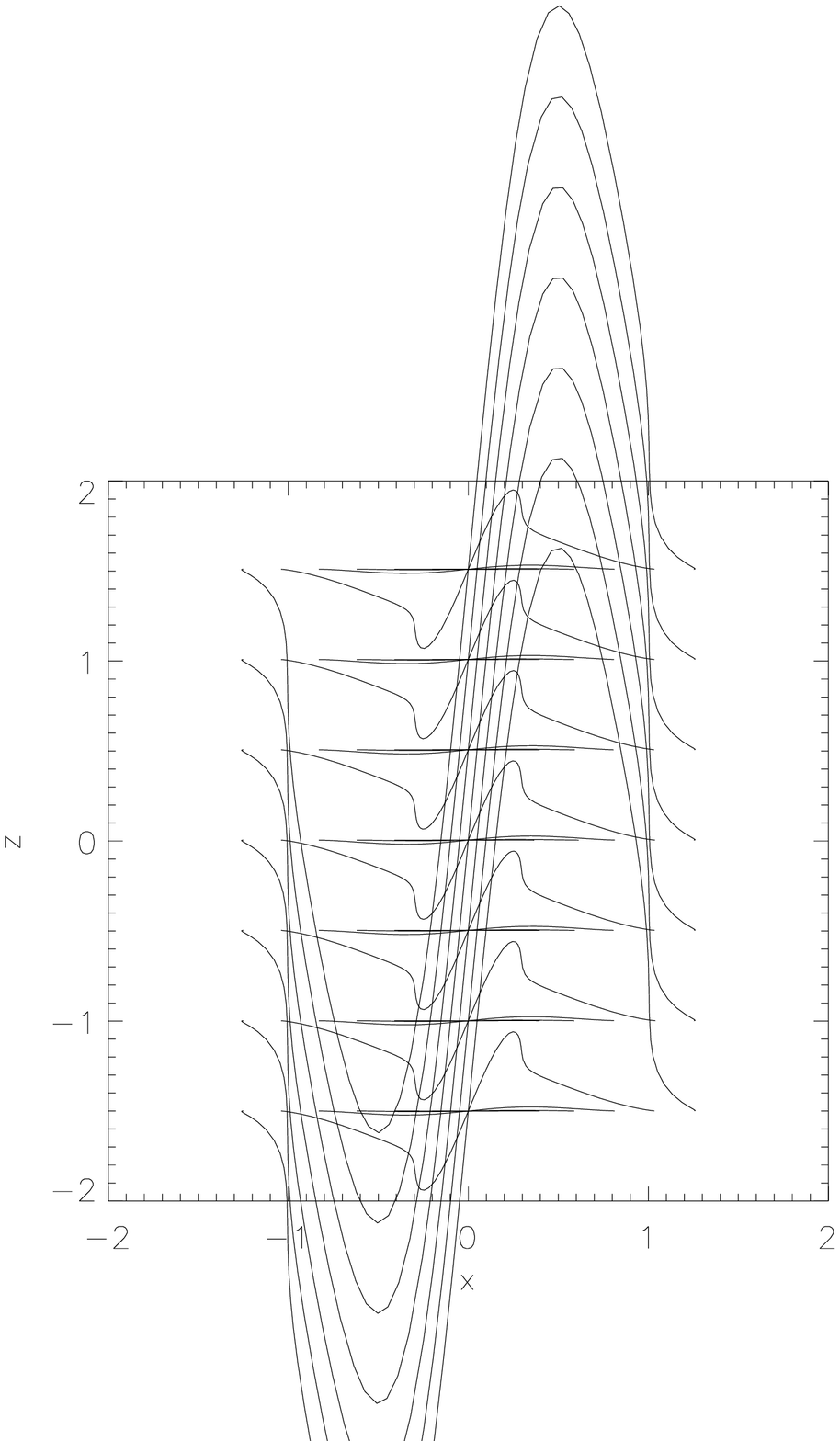}\includegraphics[clip]{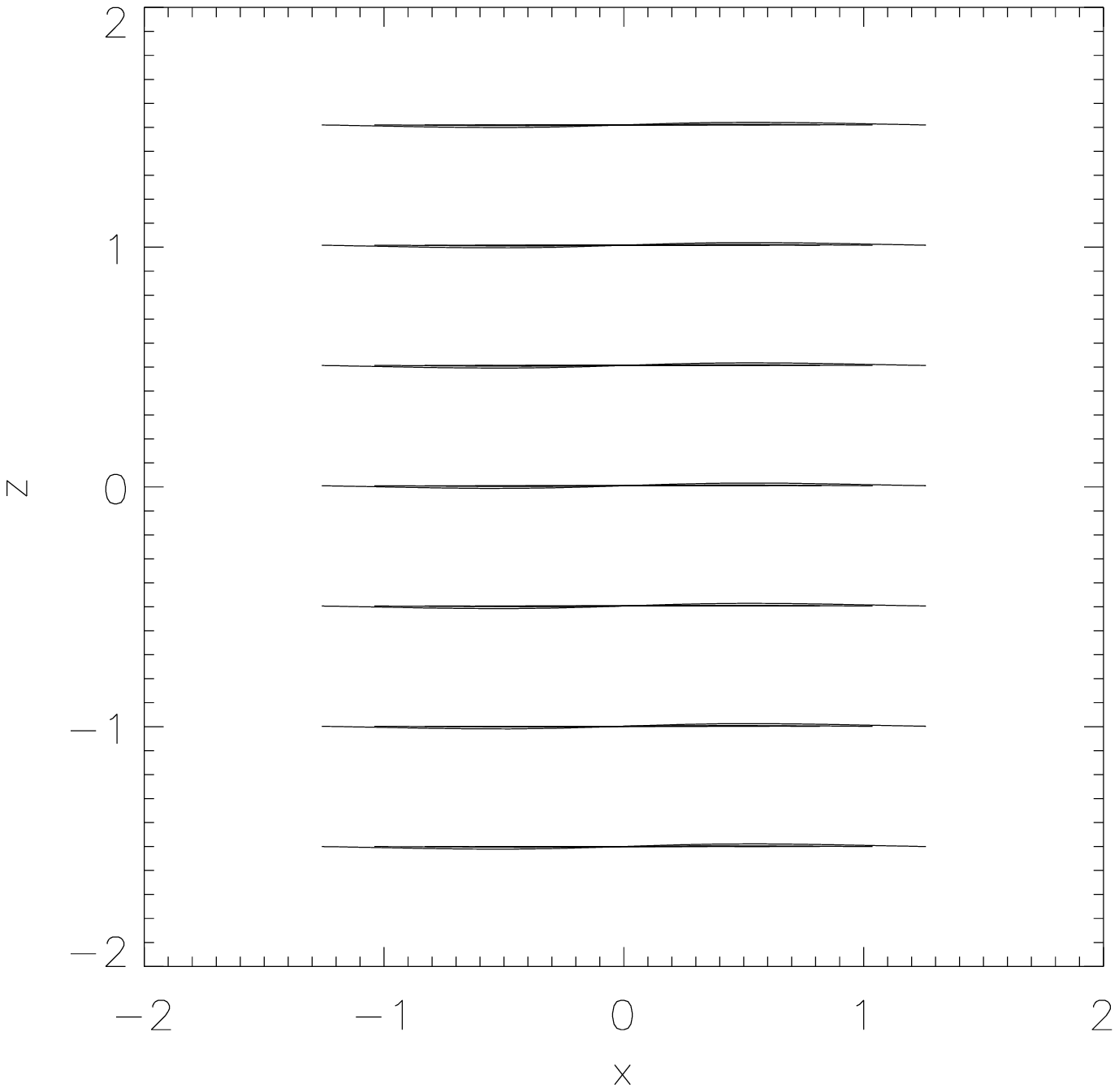}}
\caption{Top views of the field shown in Fig. \ref{fig:2d_field}, again at t=1.05s and 50.8s. These plot show more clearly how the magnetic field unshears.}
\label{fig:2d_top_field}
\end{figure}

\begin{figure}
\resizebox{\hsize}{!}{
\includegraphics[clip,bb=141 72 425 466]{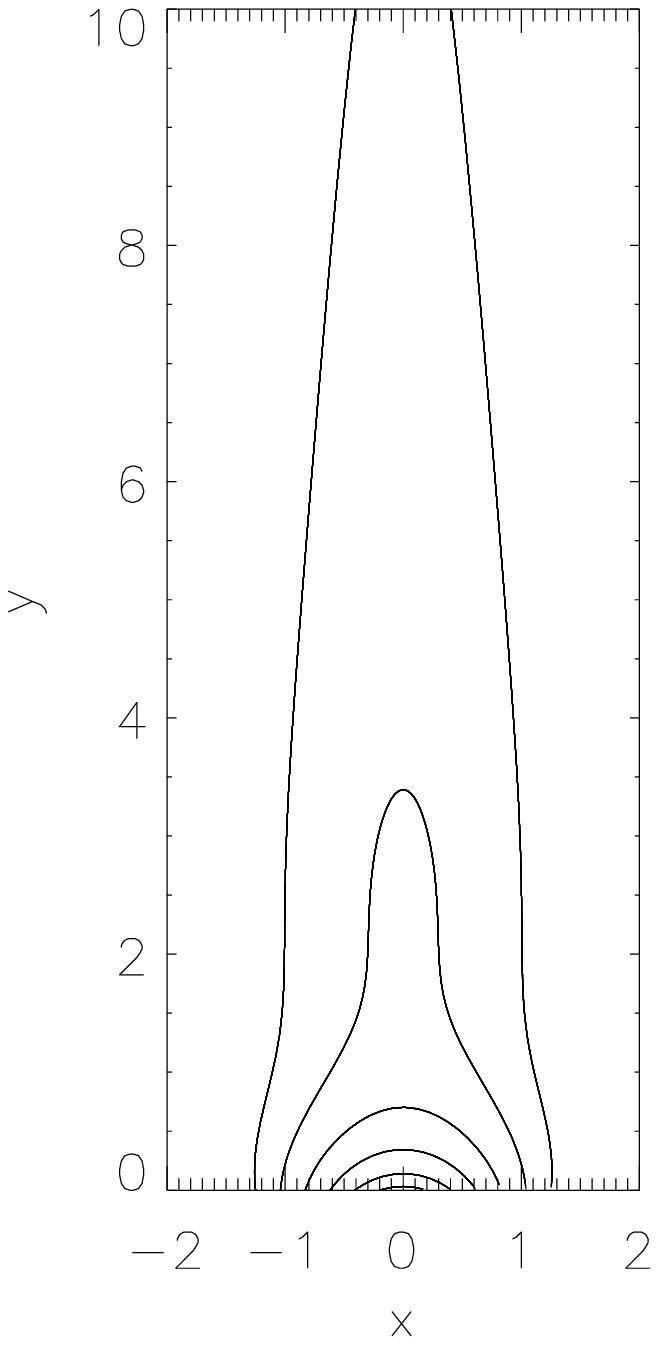}
\includegraphics[clip,bb=141 72 425 466]{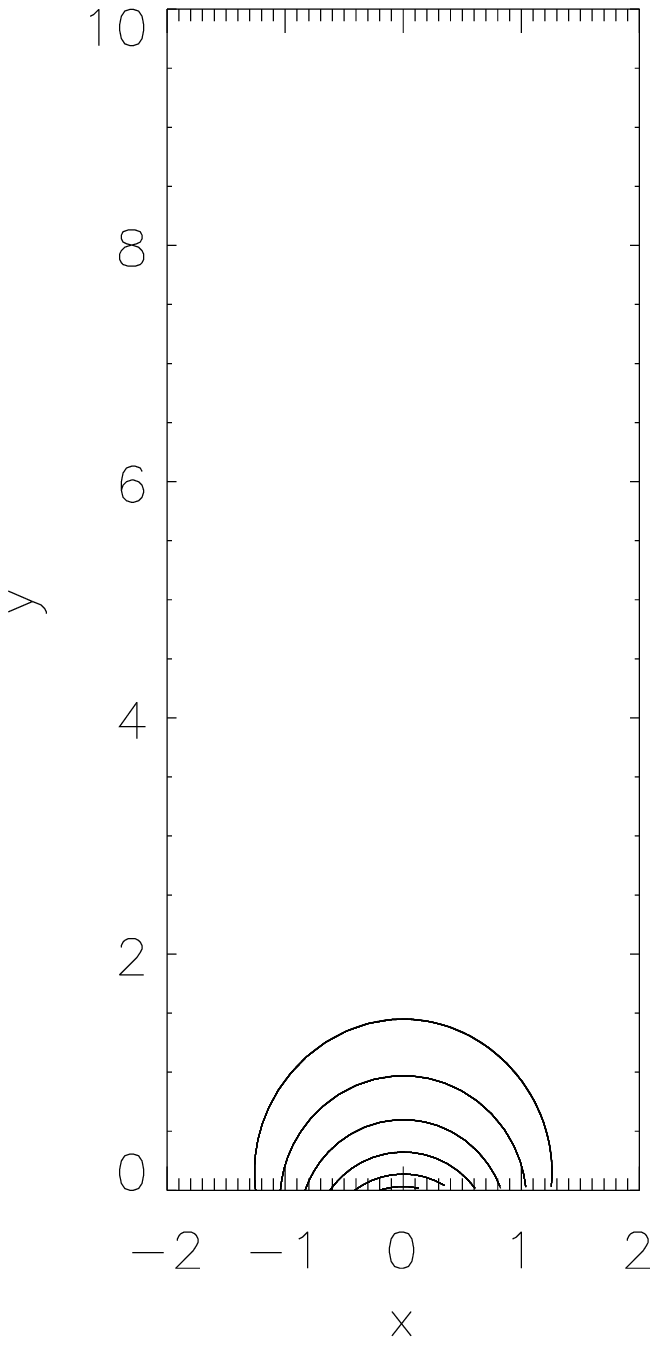}}
\caption{Side views of the field shown in Fig. \ref{fig:2d_field} at t=1.05s and 50.8s. One can clearly see
the collapse of the field lines in the CMT.}
\label{fig:2d_side_field}
\end{figure}

An example of the effect of the transformation on the magnetic field is shown in Figs. \ref{fig:2d_field} to \ref{fig:2d_side_field} for two different times (in our normalisation these are $t=1.05$s  and $t=50.8$s. In these plots the values $\delta =1$ and $a_{\rm{2.5D}}=1$ have been used. The initial shear and stretching as well as the collapse and unshearing of the field are obvious when comparing the plots of the
magnetic field for the two different times.

\subsection{An illustrative example for a 3D CMT model}
\label{sec:3d}

To generate an example model for a 3D CMT,
we use two magnetic point sources placed at positions $(-L/2,-d,0)$ and $(L/2,-d,0)$, so the sources are
located underneath the $x$-axis at depth $-d$ under the photosphere ($y=0$), and they are separated by a distance $L$. The potential magnetic field generated by these sources is then given by
\begin{eqnarray}
\vec{B_0}= & & c_1 \frac{ \left[ \left(x_0 +\frac{L}{2} \right) \mathbf{e}_x + \left(y_0 + d \right) \mathbf{e}_y +
 z_0 \; \mathbf{e}_z\right]}{\left[(x_0+L/2)^2+(y_0+d)^2+z_0^2\right]^{(3/2)}}
   \nonumber \\
&&- c_1
 \frac{\left[ \left(x_0 -\frac{L}{2} \right) \mathbf{e}_x + \left(y_0 + d \right) \mathbf{e}_y +
 z_0 \; \mathbf{e}_z\right]}{\left[(x_0-L/2)^2+(y_0+d)^2+z_0^2\right]^{(3/2)}}
 \mbox{.}
\label{3DB}
\end{eqnarray}
The value of $c_1$ is chosen so that the maximum value of the magnetic field on the photosphere is around $0.01$ T ($100$ G). We choose $c_1$ to be negative so that the magnetic polarity is negative for $x_0$ positive. As in the previous 2D case, we use $d=L$ and our standard 
normalisation $L=10^7$ m. As in the 2D case with shear flow, this potential field
is the final field to which the CMT relaxes as $t \to \infty$. It can be considered as a 3D generalisation
of the 2D magnetic field used by \citet{giuliani:etal05} and in the present paper in Sect. \ref{sec:2d}.

For this 3D example we choose a transformation which initially twists the field lines around the $y$-axis
above a given height and for a given distance from the $y$-axis, as well as stretching them in the $y$-direction as in \citet{giuliani:etal05}. The time-dependence of the transformation then untwists the field while it relaxes. To achieve this feature we now transform the $x$-coordinate as well as the $z$-coordinate, while keeping the transformation for $y$ as given in Eq. (\ref{ytransform2d}) to 
make this illustrative example more easily comparable to the work by \citet{giuliani:etal05} and the
2.5D case with shear flow described above.

The general structure of the $x$- and $z$-transformations is similar to the 2.5D case, with the difference
that the $x$-transformation now also depends on $z$, while the $z$-transformation depends on $x$ as follows:
\begin{eqnarray}
x_0=x-\delta \left( y_0(y,t) -y \right) \frac{z}{a^2_{\rm{3D}}+x^2+z^2} \mbox{,} \\
z_0=z+\delta \left( y_0(y,t) -y \right) \frac{x}{a^2_{\rm{3D}}+x^2+z^2} \mbox{.}
\end{eqnarray}
The $y$-dependence has the same effect as for the 2.5D case with shear flow, 
whereas parameters $\delta$ and  $a_{\rm{3D}}$ control the amount of twist and the distance from the $y$-axis for which there is twisting. The form of the transformation ensures that field lines which pass through the region where the transformation deviates noticeably from the identity transformation are twisted in the counterclockwise direction apart from being stretched in the $y$ direction.

\begin{figure}
\resizebox{\hsize}{!}{\includegraphics[clip]{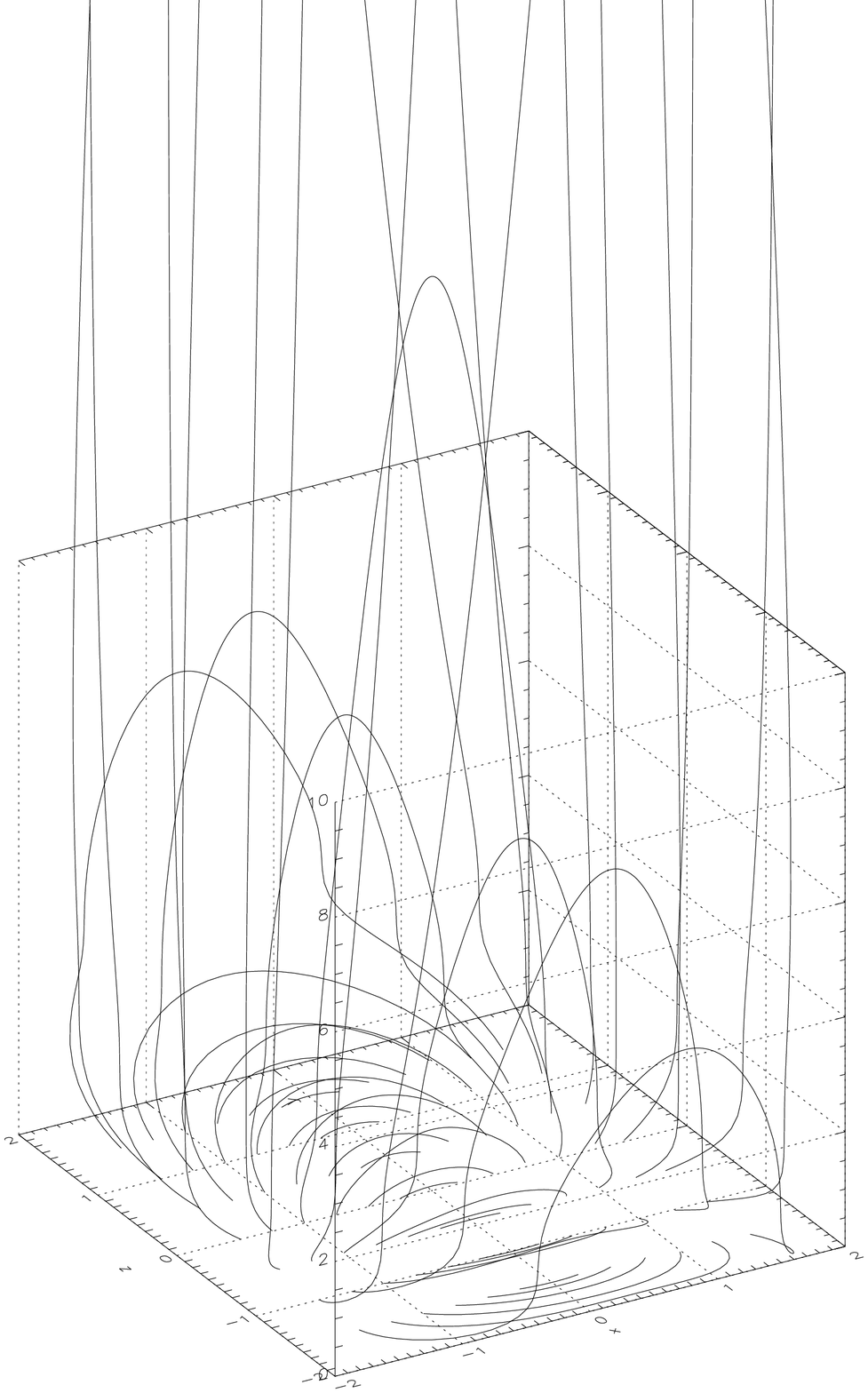}\includegraphics[clip]{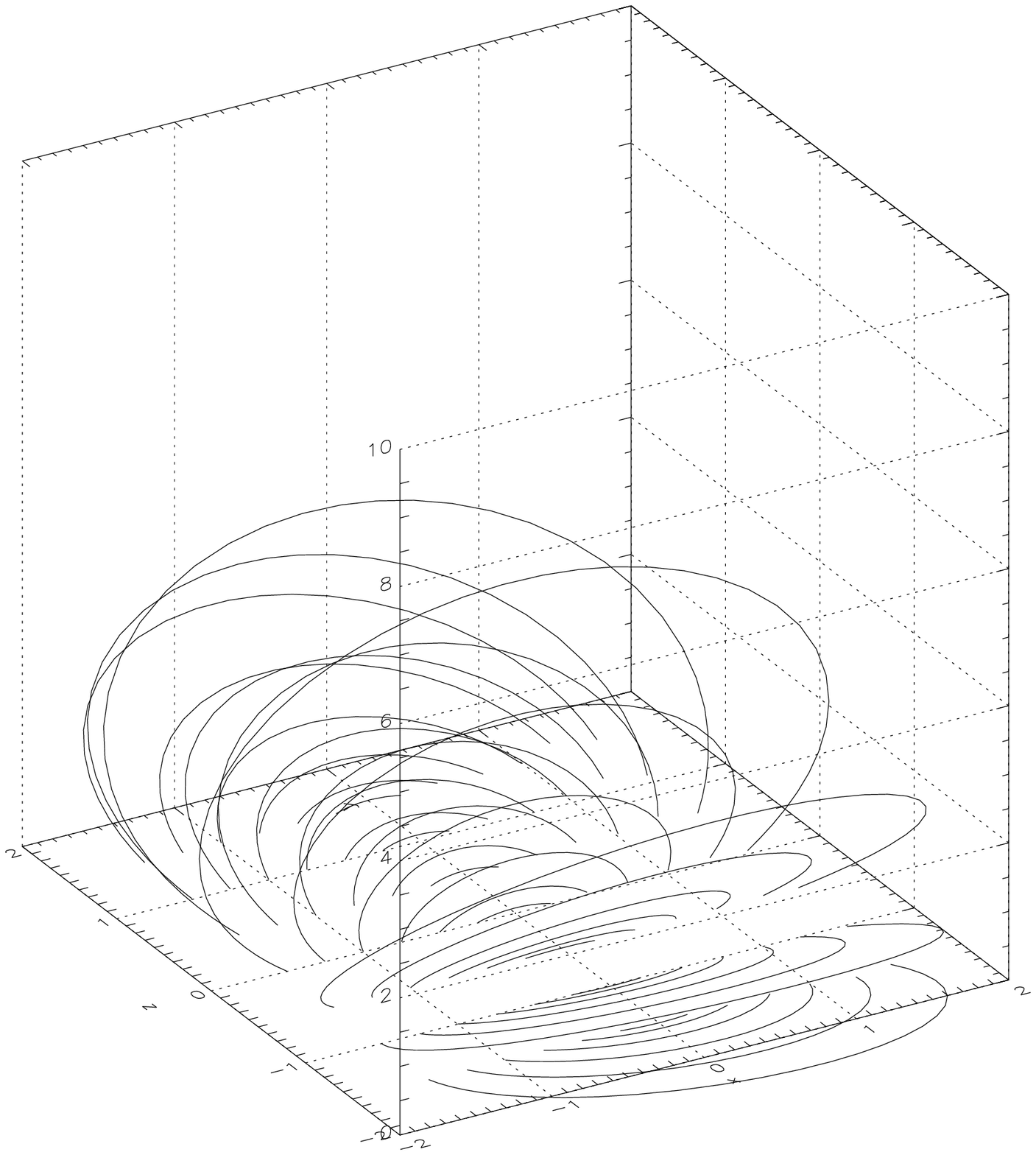}}
\caption{Field lines for the 3D example case. Lengths are normalised to $L=10Mm$. Left image shows the trap at $1.05s$, right shows once it has collapsed at $50.8s$. We point out that there is difference in scale between the $y$-direction, extending from $0$ to $10$ L in the plot and the $x$-$z$-plane which extends between $-2$ L and $2$ L in both directions.
}
\label{fig:3d_field}
\end{figure}

\begin{figure}
\resizebox{\hsize}{!}{\includegraphics[clip]{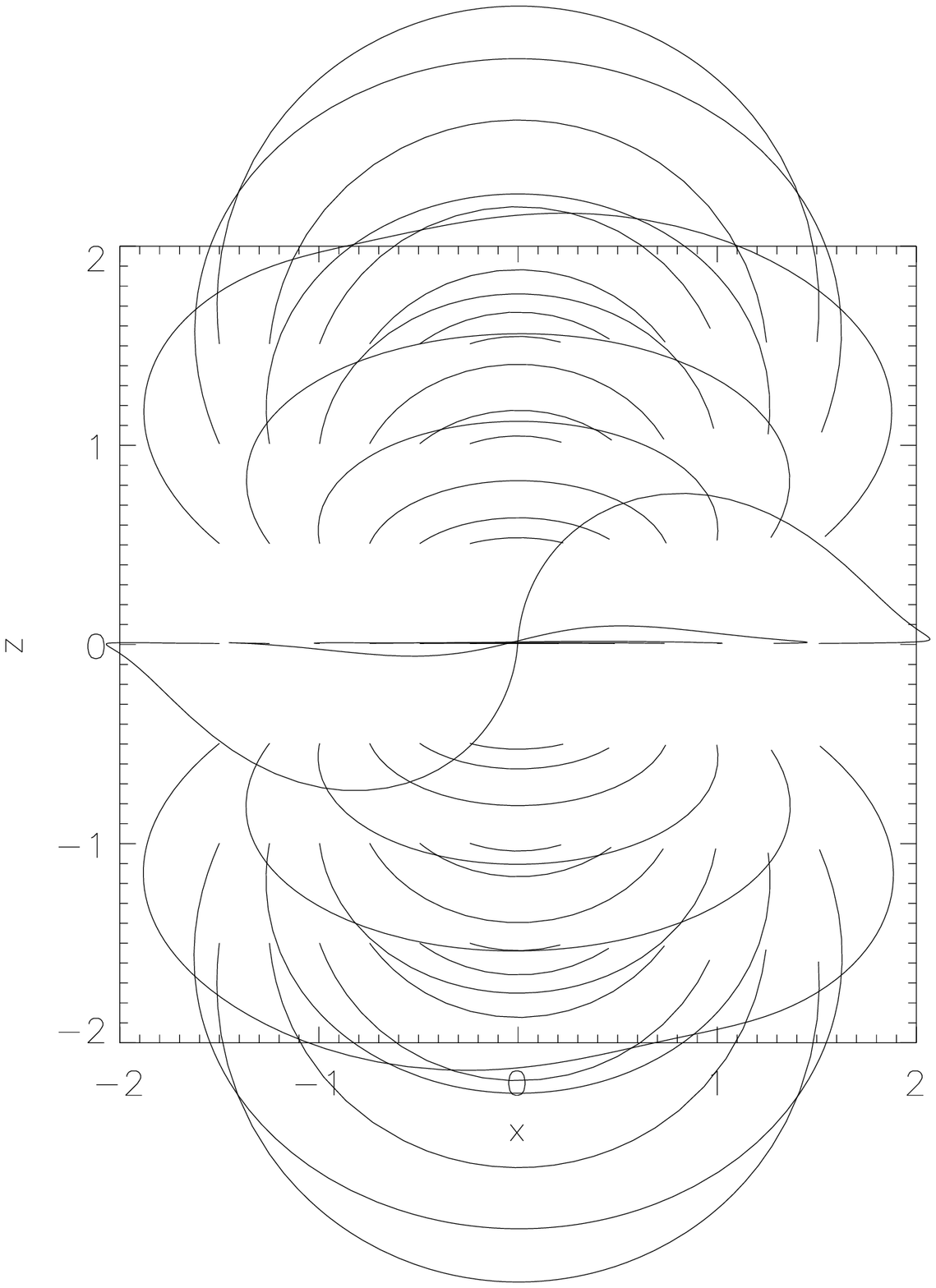}\includegraphics[clip]{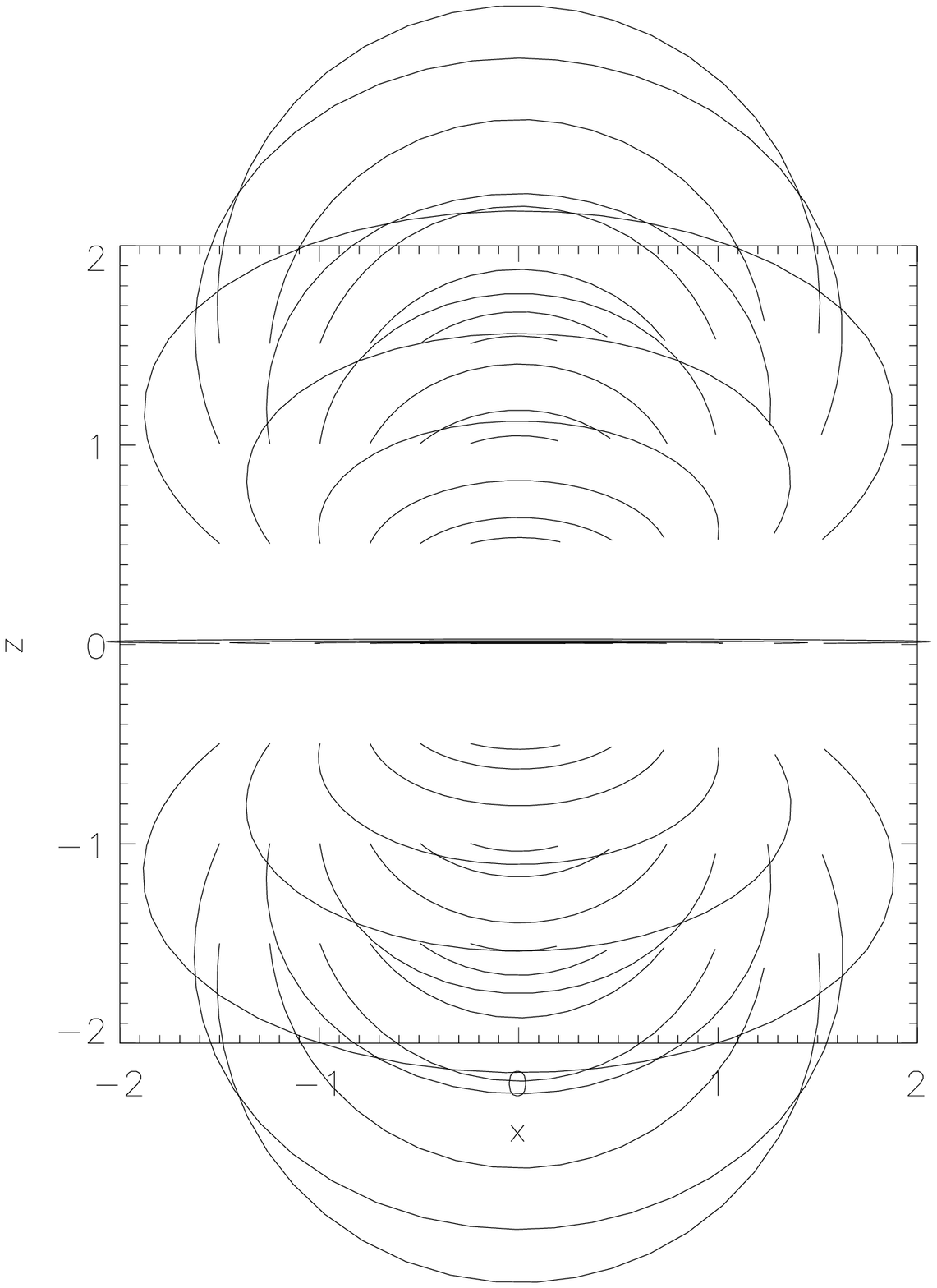}}
\caption{Top views of the field shown in Fig. \ref{fig:3d_field}, again at t=1.05s and 50.8s. These plot show more clearly how the magnetic field unshears.}
\label{fig:3d_top_field}
\end{figure}

\begin{figure}
\resizebox{\hsize}{!}{\includegraphics[clip,bb=141 72 425 466]{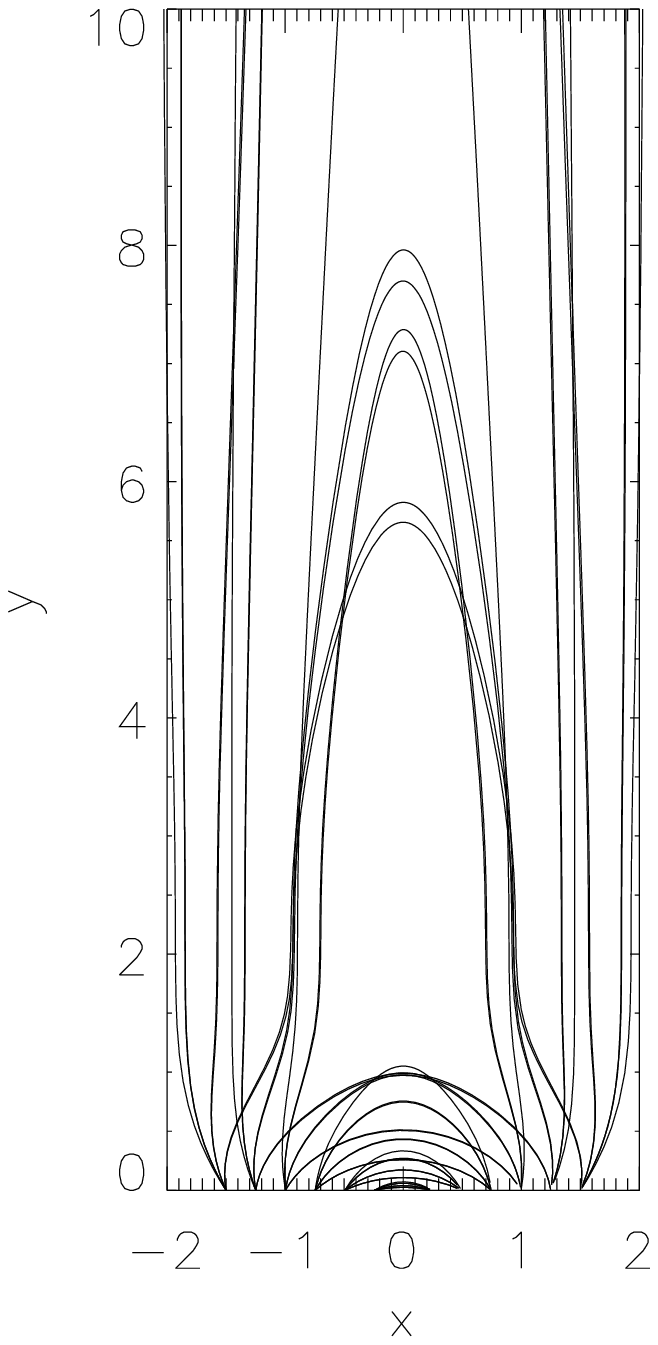}\includegraphics[clip,bb=141 72 425 466]{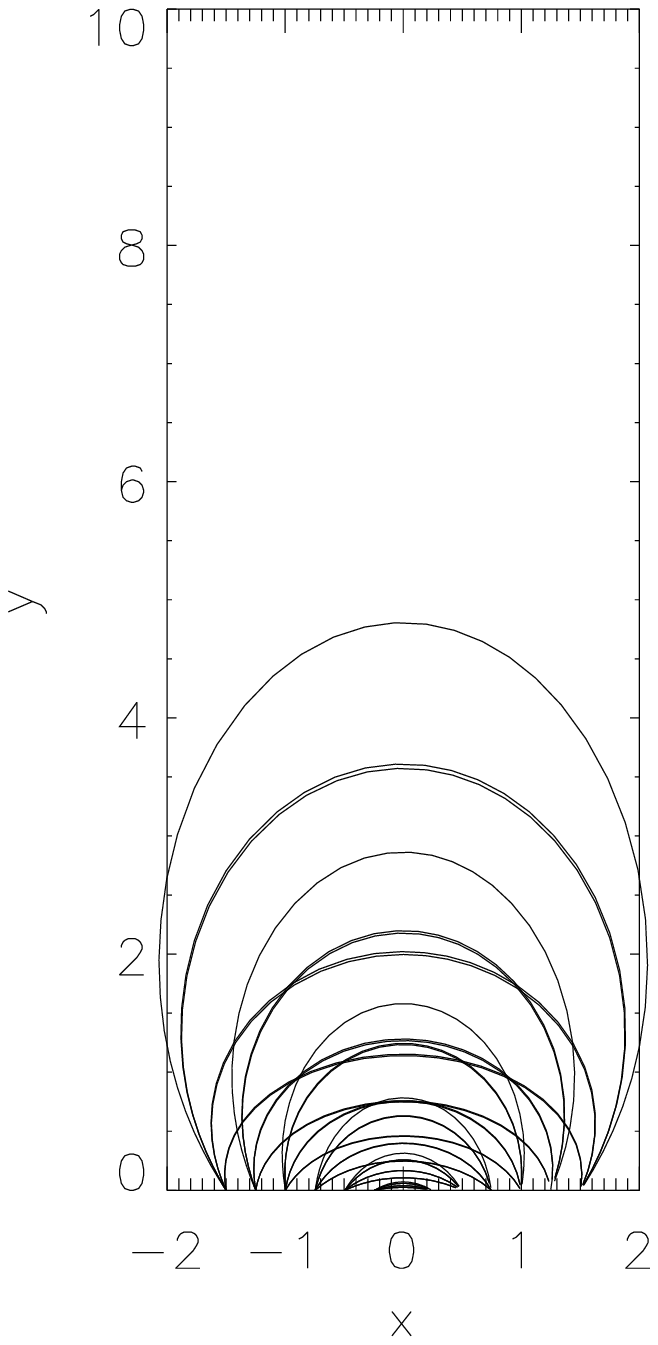}}
\caption{Side views of the field shown in Fig. \ref{fig:3d_field}, again at t=1.05s and 50.8s. The collapse of
the magnetic field lines in the CMT model is obvious.}
\label{fig:3d_side_field}
\end{figure}

We show an example with parameter values of  $\delta = 0.001$ and $a_{\rm{3D}} = 1$ in Figs. \ref{fig:3d_field}, \ref{fig:3d_top_field} and \ref{fig:3d_side_field}. Figure \ref{fig:3d_field} shows the how field lines relax between the
 initial time ($1.05$ s in the normalisation used for this example) and a later time ($50.8$ s).  
Apart from the collapse built into the example by the $y$-transformation (see Fig. \ref{fig:3d_side_field}) we can clearly see the effect of 
field line twisting through the $x$- and $z$-transformations, in particular in Fig. \ref{fig:3d_top_field}.

\section{Test Particle Orbits}
\label{sec:orbits}

We present a couple of example calculations of particle orbits for the CMT models presented above to compare them to the case studied in \citet{giuliani:etal05}.
As the gyroperiod and gyroradius of electrons are far smaller than the typical time and length scales of the collapsing traps we can use guiding centre theory to determine the particle 
trajectories \citep[e.g.][]{northrop63,giuliani:etal05}.

Initial conditions for the test particles for both the 2D example with shear flow and 3D example were chosen to be comparable to those studied in \citet{giuliani:etal05}, i.e. we have set the particles to start at the point $x=0.1$, $y=2.0$, $z=1.25 \times 10^{-6}$ in normalised coordinates. For the 2D case the 
$z$-value is of course irrelevant due to the invariance in the $z$-direction, but we choose it to be small, but non-zero, for the 3D case to avoid creating a non-generic orbit.

The value for the magnetic moment was also kept the same as in \citet{giuliani:etal05}. Because the magnetic fields at the starting positions are now different, keeping the magnetic moment the same means the initial energy of the particles is different to the $6.5$keV used by \citet{giuliani:etal05}. The values of the new magnetic fields at this starting point do not differ significantly, so the initial energies are of a similar magnitude to the previous work.

\begin{figure}
\resizebox{\hsize}{!}{\includegraphics[clip]{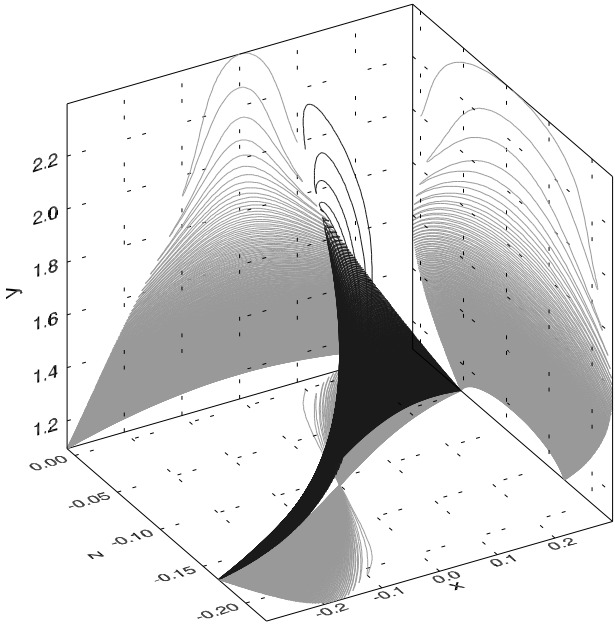}}
\caption{Particle orbit in the 2D CMT with shear flow. Projections of the trajectory onto the coordinate planes are shown on the sides of the box.}
\label{fig:2d_orbit}
\end{figure}

Figure \ref{fig:2d_orbit} shows the particle orbit for an electron in the 2D fields with shear flow. The particle follows the untwisting fieldlines, and this can be seen clearly in the projections of the trajectory onto the coordinate planes, which are shown on the sides of the box. The orbit looks otherwise similar to the 2D case without shearing as examined by \citet{giuliani:etal05}. 

\begin{figure}
\resizebox{\hsize}{!}{\includegraphics[clip]{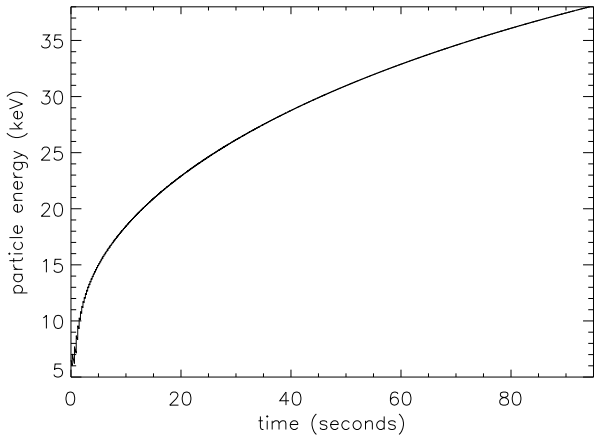}}
\caption{Time evolution of the particle energy in the 2D CMT with shear flow. This evolution is very similar to the orbit discussed by \citet{giuliani:etal05}.}
\label{fig:2d_energy}
\end{figure}

The kinetic energy of the particle as it travels through the trap is shown in Fig. \ref{fig:2d_energy}. As in the 2D case, the energy is gained initially mainly due to the effects of the curvature drift, whereas in later stages the betatron effect is stronger. The particle starts with an energy of 6.5 keV. After 95 seconds the particle energy has increased by a factor of about 6 to 38.0 keV. This is a similar gain to that seen by \citet{giuliani:etal05} using the stretched field without shear flow to accelerate an electron with initial energy of 6.5keV to 37.3 keV.

\begin{figure}
\resizebox{\hsize}{!}{\includegraphics[clip]{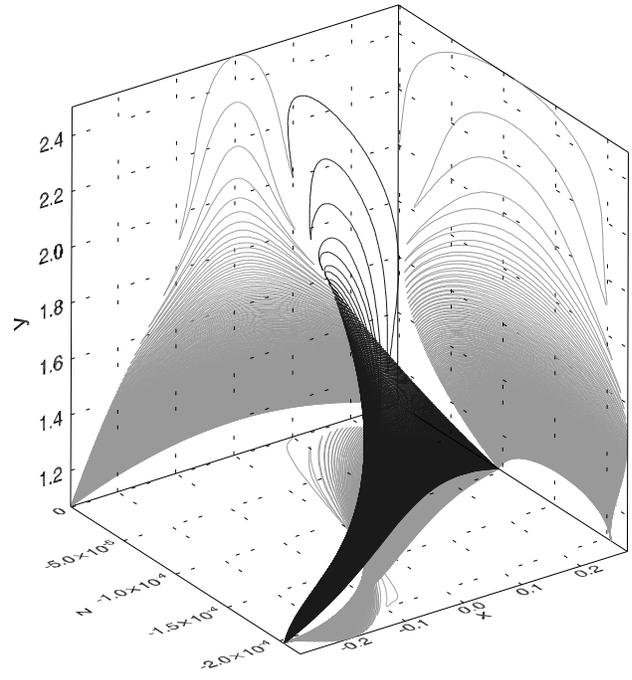}}
\caption{Particle orbit in the 3D CMT model.}
\label{fig:3d_orbit}
\end{figure}

The particle orbit in the 3D collapsing trap is shown in Fig. \ref{fig:3d_orbit}. This shows the effect of the untwisting fieldlines on the particle trajectory. A notable difference from the 2D CMT with shear flow is the asymmetric projection of the orbit onto the $x$-$z$- and $y$-$z$-planes, whereas in the $x$-$y$-plane the orbit looks very similar to the orbit in \citet{giuliani:etal05}.

\begin{figure}
\resizebox{\hsize}{!}{\includegraphics[clip]{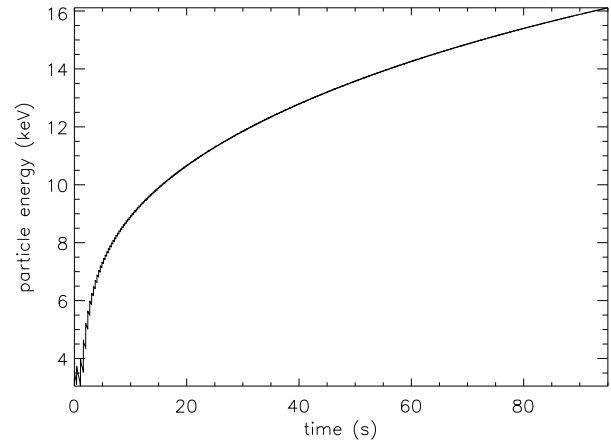}}
\caption{Energy gain of the particle in the 3D CMT model.}
\label{fig:3d_energy}
\end{figure}

The energy of the electron in the 3D example is plotted in Fig. \ref{fig:3d_energy}. As the 3D magnetic field decreases faster with height than the 2D field, the initial particle energy is lower for the same magnetic moment than in the 2D case with shear flow. The initial energy is about  3 keV and increases to about 16 keV after 95 seconds, which corresponds to an increase by a factor 5, whereas in the 2D cases we had an increase by a factor of just short of 6.

We remark that for both examples presented here, we have not yet tried to find initial conditions which give rise to higher energy gains than rather modest ones that we have found in our examples. A more systematic investigation of the 2D CMT of \citet{giuliani:etal05} shows that much larger energy gains are possible with increases of a factor 50 or more \citep{grady:etal09}. We expect similar energy gains to be possible for the cases presented here. Another reason for the rather modest increase in energy is that we have been conservative in our assumptions about the maximum magnetic field strength on the photosphere, which is only about 100 G. A factor 5 to 10 increase of the photospheric field strength seems reasonable, in particular for flaring regions, and this could have a significant effect on energy gain. We plan to investigate this in the future.

\section{Summary and Conclusions}
\label{sec:summary}

We have developed a fully analytical model for kinematic time-dependent, 2D and 3D collapsing magnetic traps. This kinematic approach has the advantage that it allows us full control over all the features of the model, but has the disadvantage the modelling of the plasma system is not self-consistent. 

In the present paper, we have shown how to build kinematic CMT models using the magnetic field directly, rather than using a flux function or Euler potentials. This is much easier and more straightforward to use, especially in 3D, than the theory presented in \citet{giuliani:etal05}.

We have given illustrative examples of collapsing traps with transformations that give rise a shear flow in 2D and magnetic twist in 3D. We have calculated particle orbits for these new CMT models using guiding centre theory.
For those orbits, the CMT models were found to give similar relative energy gains as with the 2D CMT without shearing. The particle orbits are different from the 2D CMT model by \citet{giuliani:etal05} despite starting from the same initial position due to the differences in field line motion caused by the shear flow in 2D and by the twisting motion in 3D.
The examples shown in this paper have been chosen specifically to be comparable with the example shown in \citet{giuliani:etal05}. Different CMT models could allow for higher energy gains and  will be considered in future work.
There are also many other possible combinations of initial positions, initial particle energy and pitch angles, as well as investigating proton/ion orbits as well as electron orbits. 
A systematic investigation for the 2D model of \citet{giuliani:etal05} has shown that energy gain factors of order 50 or higher are possible for that model \citep{grady:etal09}.
A similar investigation is planned for the future for 2D with shear flow CMTs and 3D CMTs using the theory presented in this paper.

\begin{acknowledgements}
We thank the anonymous referee for useful comments.
The authors acknowledge financial support by the UK's Science and Technology Facilities Council and by the European Commission through the SOLAIRE Network (MTRN-CT-2006-035484).
\end{acknowledgements}

\bibliographystyle{aa}
\bibliography{references_tn}

\appendix
\section{\label{appendixa}Detailed calculation for the 3D case using Euler potentials}

For the following derivation we use a notation which allows us to handle as vectors certain groups of scalar quantities or rows or columns of tensors. Firstly, the derivatives of the Clebsch variables, 
$\bar{\alpha}$ and $\bar{\beta}$, with respect to the transformed coordinates are required:
\begin{eqnarray*}
 \frac{\partial \bar{\alpha}}{\partial{}\vec{X}} & = &
   \left(\frac{\partial \bar{\alpha}}{\partial{}X},\frac{\partial \bar{\alpha}}{\partial{}Y},\frac{\partial \bar{\alpha}}{\partial{}Z} \right) \mbox{,} \\
 \frac{\partial \bar{\beta}}{\partial{}\vec{X}} & = &
   \left(\frac{\partial \bar{\beta}}{\partial{}X},\frac{\partial \bar{\beta}}{\partial{}Y},\frac{\partial \bar{\beta}}{\partial{}Z} \right) \mbox{,}
\end{eqnarray*}
which is basically the usual gradient with respect to $X$, $Y$ and $Z$.
We also need the transformation differentiated with respect to the original Eulerian coordinates.
\begin{eqnarray*}
 \frac{\partial{}\vec{X}}{\partial{}x} & = &
   \left(\frac{\partial{}X}{\partial{}x},\frac{\partial{}Y}{\partial{}x},\frac{\partial{}Z}{\partial{}x} \right) \mbox{,}\\
 \frac{\partial{}\vec{X}}{\partial{}y} & = &
   \left(\frac{\partial{}X}{\partial{}y},\frac{\partial{}Y}{\partial{}y},\frac{\partial{}Z}{\partial{}y} \right) \mbox{,} \\
 \frac{\partial{}\vec{X}}{\partial{}z} & = &
   \left(\frac{\partial{}X}{\partial{}z},\frac{\partial{}Y}{\partial{}z},\frac{\partial{}Z}{\partial{}z} \right) \mbox{.}
\end{eqnarray*}
We now consider each component of Eq. (\ref{eq:euler}), starting with the $x$-component
\begin{equation}
  B_x= \frac{\partial \alpha}{\partial y} \frac{\partial \beta}{\partial z} -
    \frac{\partial \alpha}{\partial z} \frac{\partial \beta}{\partial y} \mbox{.}
\end{equation}
With the coordinate transformation, Eqs. \ref{eq:solalpha} and (\ref{eq:solbeta}),
and using the chain rule this becomes
\begin{equation}
      B_x = \left( \frac{\partial \bar{\alpha}}{\partial \vec{X}} \cdot \frac{\partial \vec{X}}{\partial y} \right)
            \left( \frac{\partial \bar{\beta}}{\partial \vec{X}} \cdot \frac{\partial \vec{X}}{\partial z} \right)
       -  \left( \frac{\partial \bar{\alpha}}{\partial \vec{X}} \cdot \frac{\partial \vec{X}}{\partial z} \right)
            \left( \frac{\partial \bar{\beta}}{\partial \vec{X}} \cdot \frac{\partial \vec{X}}{\partial y} \right)\label{eq:beforeapp} \mbox{.}
\end{equation}
Applying the well-known vector identity
\begin{equation}
(\vec{A}\cdot\vec{C})(\vec{B}\cdot\vec{D}) - (\vec{A}\cdot\vec{D}) (\vec{B}\cdot\vec{C}) =
(\vec{A}\times\vec{B}) \cdot(\vec{C}\times\vec{D})
\end{equation}
to Eq. (\ref{eq:beforeapp}) we arrive at
\begin{equation}
      B_x  = \left( \frac{\partial{}\vec{X}}{\partial{}y} \times \frac{\partial{}\vec{X}}{\partial{}z} \right) \cdot
           \vec{B}_0 \left( \vec{X} \right) \mbox{,}
           \label{eq:start_of_3d} 
\end{equation}
because the initial magnetic field is
\begin{equation}
  \vec{B}_{0}\!\left(\vec{X}\right) = \frac{\partial \bar{\alpha}}{\partial{}\vec{X}} \times \frac{\partial \bar{\beta}}{\partial{}\vec{X}} \label{eq:initialB}
\end{equation}
by construction.
Similarly one finds that 
\begin{eqnarray}
      B_y &=&  \left( \frac{\partial{}\vec{X}}{\partial z} \times \frac{\partial{}\vec{X}}{\partial x} \right) \cdot
           \vec{B}_0 \left( \vec{X} \right), \\
      B_z  &=& \left( \frac{\partial{}\vec{X}}{\partial x} \times \frac{\partial{}\vec{X}}{\partial{}y} \right) \cdot
           \vec{B}_0 \left( \vec{X} \right).
\end{eqnarray}

\end{document}